\documentclass[preprint2]{aastex}
\usepackage{amsmath}

\newcommand{\kms}{km~s$^{-1}$}

\newcommand{\etal}{{\em et al.\/}}
\newcommand{\asa}{{\it Astron. Astrophys.\/}}

\newcommand{\sps}{{\it Solar Phys.\/}}
\usepackage{graphicx}
\usepackage{epsfig}          

\begin{document}

\title{Combining particle acceleration and coronal heating via 
data-constrained calculations of nanoflares in coronal loops}
\author{C. Gontikakis$^1$, S. Patsourakos$^2$, C. Efthymiopoulos$^1$,
A. Anastasiadis$^3$,and M.K. Georgoulis$^1$}

\affil{$^1$Reseach Center for Astronomy and Applied Mathematics,
Academy of Athens, Soranou Efessiou 4, 11528, Athens Greece}
\email{cgontik@academyofathens.gr}
\affil{$^2$Section of Astro-Geophysics, Department of Physics,
University of Ioannina, 45110 Ioannina, Greece}
\affil{$^3$National Observatory of Athens, Institute for Astronomy, Astrophysics, Space Applications and Remote Sensing, GR-15236, Palaia Penteli, Greece}

\begin{abstract}
We model nanoflare heating of extrapolated active-region coronal loops
via the acceleration of electrons and protons in Harris-type
current sheets. The kinetic energy of the accelerated particles
is estimated using semi-analytical and test-particle-tracing
approaches. Vector magnetograms and photospheric Doppler velocity maps
of NOAA active region 09114, recorded by the Imaging Vector
Magnetograph  (IVM), were used for this analysis. A current-free field
extrapolation of the active-region corona was first constructed. 
The corresponding Poynting fluxes at the footpoints of 5000
extrapolated coronal loops were then calculated. Assuming that 
reconnecting current sheets develop along these loops, we utilized
previous results to estimate the kinetic-energy gain of the accelerated 
particles and we related this energy to nanoflare heating and macroscopic loop characteristics.
Kinetic energies of 0.1 to 8~keV (for electrons) and
0.3 to 470~keV (for protons) were found to cause heating rates
ranging from $10^{-6}$ to 1 $\mathrm{erg\, s^{-1} cm^{-3}}$.
Hydrodynamic simulations show that such heating rates can
sustain plasma in coronal conditions inside the loops and generate plasma thermal distributions which are consistent with active region observations.
We concluded the analysis by computing the form of X-ray spectra generated by the
accelerated electrons using the thick target approach that were found to be 
in agreement with observed X-ray spectra, thus supporting the plausibility of our nanoflare-heating scenario.
\end{abstract}

\keywords{Sun: activity, Sun: corona, Sun: flares,
Sun: magnetic topology}

\section{Introduction}
A strong candidate mechanism to explain coronal heating is the
formation of small-scale, still undetected, reconnection events,
called nanoflares \citep{Par88}. Reconnection events are sites where 
coronal magnetic energy is transformed into energy of accelerated particles 
eventually producing heating.
Solar flares constitute a well documented case where accelerated particles, 
through reconnection, attain higher and higher kinetic energies,
thus raising the plasma temperature while creating in the same
time particle beams of supra-thermal kinetic energies \citep{Birn07}.\\

Particle acceleration and the thermodynamic response of the
plasma to heating are phenomena that were typically studied 
in isolation. In fact, a unifying study of particle
acceleration, direct coronal heating and the thermodynamic
response of the observed plasma structures such as coronal loops,
seems hardly tractable, given the wide range of spatial
and/or temporal scales involved.
Indeed, these scales range from the dissipation scale
(i.e. the scale of current sheet thickness), which is of the
order of centimeters or meters, to the scale of macroscopic phenomena
taking place in the observed coronal structures which is of the order of several tens to $\simeq 100$~Mm \citep{Klim06}. \\

In the present work, we report on the results of an attempt to connect direct coronal heating, and its thermodynamic 
response, to particle acceleration. For this we use several simplifying assumptions in the theoretical
and numerical treatment of
each individual process considered. Furthermore, we empirically
constrain our assumptions by exploiting available observational information, 
namely, measurements and estimates of the magnetic and the velocity field vectors in a particular active-region.

As a starting point we adopt Parker's hypothesis
\citep{Par72} that plasma motions at photospheric level
stress, twist, and entangle the coronal magnetic field lines. This process
converts the plasma kinetic-energy at the
footpoints of coronal loops to non-potential (free) magnetic energy stored in the
coronal magnetic fields. According to \citet{Par88}, when the
magnetic field stress reaches a critical point, the stored free magnetic
energy should be released to the plasma via reconnection events.
For quantitative calculations, one equates the work rate done by the photospheric
motions to the observed radiated energy in the corona, thus finding
an estimate of the mean critical inclination of flux tubes, with respect to the
direction normal to the solar surface, at which a reconnection event should take place. The resulting inclination
angle, called the Parker angle, is derived through numerical simulations 
or analytical estimations. For active regions, it takes values 
in the range of 5$\degr$ to 20$\degr$ depending on the applied
physical mechanism \citet{Gal96,Klim06,Rap07}. The reason 
why the Parker angle appears preferentially in the above range (instead
of taking values below 1$\degr$ or around $90\degr$) has been interpreted in
some numerical simulations as an effect of the temporal evolution of
a reconnecting current sheet undergoing tearing instability
\citep{Dahl05}.

Numerical simulations can also help study how coronal magnetic fields
are stretched and twisted due to photospheric plasma motions, thus
developing complex, unstable current sheets 
(see, for example, \citet{Gal96}). In some cases it was found that there is
a statistical equilibrium established between the energy supplied at
the loop footpoints and the energy released throughout the whole
structure of coronal loops via current sheets \citep{Gal96,Hen96}.
The simulated current sheets are of various types,
scales, and forms. In particular, they vary from large monolithic
structures extending over the loop's length \citep{Gal96} to
cascades of smaller structures exhibiting a range of different sizes and
spatial distributions \citep{Gal96,Hen96,Rap07}. 

The above simulations show the dependence of the Parker angle on 
the loop parameters. However, they still present a high degree 
of idealization which may affect the computed Parker angle values. These 
include the simplified representation of the region between the photosphere and the corona 
\citep{Klim06}, as well as the omission of the flux-tube divergence with height.

As our basic mechanism to explain the magnetic energy release during
flares and nanoflares we adopt the acceleration of particles inside
reconnecting current sheets. A crucial parameter toward achieving  a 
viable simulation of this type is the \lq guide\rq\ magnetic field, i.e. a magnetic
field component parallel to the electric field that accelerates the
particles. The main effect of the guide magnetic field is to change
the {\it trajectories} of the charged particles, thus enhancing
motion parallel to the electric field \citep{Litv93,Litv96,Litv00}.
In a previous work \citep{Efth05}, these particles' motions were studied
for the case of a Harris-type reconnecting current sheet model by means of
a Hamiltonian formalism. The result was a general formula predicting
the maximum kinetic-energy gain of accelerated particles
as a function of the initial energy of the particles and the parameters 
of the current sheet (thickness; field strengths). This formula
is used in the present work to estimate the kinetic energy of the
accelerated particles in the extrapolated coronal loops. We also note that, as found
in \citet{Gont07} and \citet{Ana08}, the kinetic-energy distributions
of accelerated particles through single or multiple current sheets
(Harris type or X-point) are subject to upper limits due to the existence of a maximum possible
kinetic energy gain of the particles.

Another input of our present analysis are estimates of the current sheet
thicknesses encountered in these reconnection events. For this, we
exploit results of recent particle-in-cell studies,
in which the magnetic reconnection in solar flares
takes place under the presence of a guide magnetic field \citep{Hess99,Cas08}.
One then finds that the thickness of the diffusive
current sheet is of the order of the electron
gyro-radius.

Considering now the response of the coronal plasma to 
nanoflares, we rely on time-dependent
hydrodynamic simulations \citep{Pats05,Pats04}. In these models,
the cumulative effect of a large number of nanoflares, releasing energy
in a coronal loop, is simulated in a way that allows direct comparison between simulations and
observations from telescopes
such as the {\it Transition Region and Coronal Explorer} (TRACE)
\citep{Hand98}  or, the {\it Atmospheric Imaging Assembly}
(AIA, \citet{Le12}) on board of the {\it Solar Dynamic Observatory}
(SDO).

The structure of the paper is as follows: in Section 2 we discuss
the observational data used to compute the structure of the magnetic field 
in the active-region corona. Section~3 presents our modeling
of coronal loops, leading to a derivation of values for the field
strengths and the footpoint velocities of the field lines
forming the loops. Section 4 explains our main assumptions used 
to derive needed values of the current sheet parameters.
Section~5 presents the particle acceleration results. These are used as an input 
to compute, in Section 6, the overall loop heating caused
by particle acceleration. In Section~7 we present a simulation
of how the X-Ray spectra of the accelerated electrons would look
like under a thick-target model. Section~8 contains comprehensive results
on loop heating via nanoflares, as well as on 
the thermodynamic response of the plasma in the modeled coronal loops.
Section 9 offers a discussion of our results and of the limitations and validity of our modeling. 
Finally, Section~10 summarizes the basic conclusions of the present
study.

\section{Observations and data treatment}
The Imaging Vector Magnetograph of the University of Hawaii's Mees 
Solar Observatory recorded a timeseries of 12 vector magnetograms 
of the active region (AR) 09114 spanning over a $\simeq$4.5-hour period 
on 8 August 2000 with a cadence of 20~minutes. These
magnetograms have a field of view of 280\arcsec $\times$ 280\arcsec, 
and a spatial resolution of 0.55\arcsec\ \citep{Mic96}. The IVM 
recorded the Stokes vector of the Fe~{\sc I} 6302.5~\AA\ photospheric 
spectral line. 
\begin{figure}[!h]
\epsscale{1.0}
\plotone{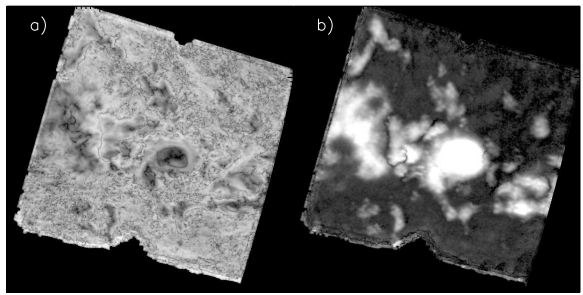}
\caption{(a) The modulus of the velocity $v=\sqrt{v_x^2+v_y^2}$ computed
with the MSR vector. The gray scale is set in the range from 0.3 to
4.5~\kms. (b) The modulus of the magnetic field from the vector
magnetogram. The gray scale is set in the range from 10 to 500~Gauss.}
\label{fig:IVMdata}
\end{figure}

Starting from the above data, we first resolve the azimuthal 
180-degree ambiguity in each magnetogram using the Non-potential 
Magnetic Field Calculation (NPFC) method of \citet{Geor05}, as refined in \citet{Met06}.
Then, we compute a mean vector magnetogram as well as a map of the 
associated photospheric horizontal velocity, calculated by means of the
Minimum Structure Reconstruction (MSR) technique \citep{Geor06} 
(see Figure~\ref{fig:IVMdata}).

Expressing the photospheric magnetic field vector in the local 
heliographic reference system, we can now perform a current-free (potential) field 
extrapolation using the method proposed by \citet{Alis81}. This 
computation was done in a cube up to an altitude of 70\arcsec, or 
$\simeq 50$~Mm, while the potential field extrapolation was performed 
on a $2\times 2$ binned magnetogram.

\section{Coronal loop modeling}
\begin{figure}[!h]
\epsscale{1.}
\plotone{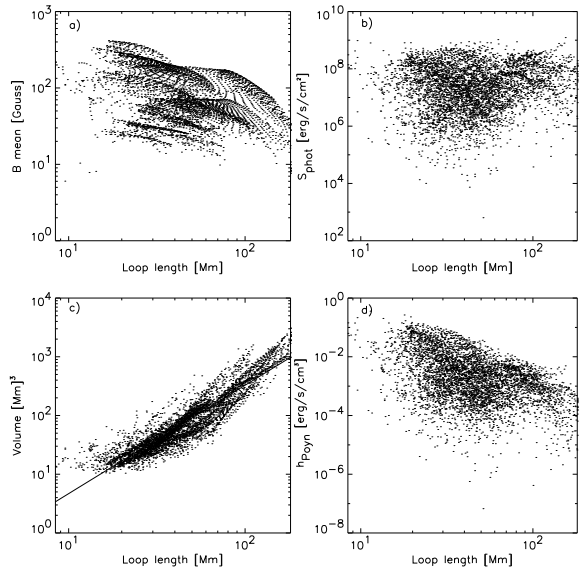}
\caption{Scatter plots for several loop parameters as a function 
of the loop length. These are (a) the mean field strength, (b) the 
Poynting flux at both footpoints of each loop, (c) the loop volume, 
and (d) the volumetric heating rate due to Poynting flux. In panel (c) a fit is also shown.}
\label{fig:Bmean}
\end{figure}

Using the extrapolated potential fields, we now define the 
coronal loops and several associated quantities.
We proceed via the following steps:

i) {\it Coronal loop identification.}
We selected about 5000 extrapolated magnetic field lines all closing within
the photospheric field of view. These magnetic field lines are identified 
as closed coronal loops. The loops' lengths (denoted $L$ hereafter) 
range from 8 to 180~Mm. Loops' maximum heights range from 2.5~Mm to 50~Mm.

ii) {\it Magnetic field strength.}
We computed a mean value $\bar B$ of the magnetic field along 
each coronal loop as in \citet{Man00,Gont08}. 
The mean magnetic field is found to statistically decrease 
with increasing loop length, with short loops ($L<20$~Mm) 
yielding an average $\bar B \simeq 180~G$, while long loops 
($L > 100$~Mm), yield an average $\bar B \simeq 60$~G. 
Figure~\ref{fig:Bmean}a shows the mean magnetic field 
strength for each loop as a function of the loop's length. 

iii){\it Electric field strength.}
For each coronal loop we calculate the photospheric electric 
field values at both footpoints $\vec E_{phot}=- \frac{1}{c} \vec v_{phot} \times \vec B$ or 
\begin{equation}\label{E_phot}
\vec E_{phot}\,=\, \frac{1}{c} ( v_y\, B_z \vec x - v_x\, B_z \vec y)
\end{equation}
In Equation~\ref{E_phot}, $v_x$ and $v_y$ are the calculated horizontal components 
of the photospheric velocities, $B_z$ is the perpendicular 
magnetic field component and $\vec x$, $\vec y$  are 
the corresponding unit vectors. Equation~(\ref{E_phot}) does not include $v_z$ because,
in the MSR method, the vertical (cross-field) velocity component is assumed to be negligible. 

iv) {\it Poynting flux through loops and supply of free 
magnetic energy}. As an immediate consequence of (iii), 
we can deduce values of the Poynting flux normal to the solar 
surface at the loops footpoints. This is expressed as
\begin{equation}\label{Poynt_phot_eq}
\vec S_{foot} = \frac{c}{4 \pi}  (E_x\, B_y - E_y\, B_x) \vec z
\end{equation}
where $E_x\,=\, \frac{1}{c}  v_y\, B_z$ and $E_y\,=\, - \frac{1}{c} v_x\, B_z$ are the 
photospheric electric field components. 
For each loop, two Poynting fluxes $\vec S_{foot1}$ and 
$\vec S_{foot2}$ are computed at the positions of the respective 
footpoints (1 or 2). We should emphasize that, while we have 
the information about the direction of the velocity vectors at 
the points of the observational grid, the local direction of 
small scale motions at the footpoints of each coronal loop is 
inaccessible. For this reason, we ignore signs indicating an 
in or out Poynting flux through the loop and define, instead, 
the absolute sum $S_{phot}=|S_{foot1}+S_{foot2}|$ as a rough 
measure of the Poynting flux supplied to the loop due to 
photospheric motions. For a single, isolated loop, the supply 
of free magnetic energy should be considered as the result of 
the relative plasma motions at both footpoints, since both distort 
the magnetic field. In fact, this distortion results in currents 
developed inside the loops, whose magnetic field now becomes non-potential.
Furthermore, Equation~(\ref{Poynt_phot_eq})
is not exact as it does not precisely resolve the relative motions at the 
footpoints. However, it arguably 
gives the correct order of magnitude for the Poynting flux. 
Finally, Equation~(\ref{Poynt_phot_eq}) neglects the 
changes of the magnetic field due to the emergence or submergence of 
magnetic flux (Harra \etal\ 2011). This is justified in the case 
of the AR 09114, since this is a fully developed and non-decaying active
region at the time of the observations.

The calculated $S_{phot}$ (Figure~\ref{fig:Bmean}b) has an average 
of $7 \times 10^7$~$\mathrm{erg\, s^{-1}\, cm^{-2}}$ and a standard deviation of 
$\simeq\ 10^8$~$\mathrm{erg\, s^{-1}\, cm^{-2}}$. 95\% of its calculated 
values are in the range $5\times 10^5$ to $5 \times 10^8$~$\mathrm{erg\, s^{-1} cm^{-2}}$, 
regardless of the loop length. In Figure~\ref{fig:Bmean}d we show 
the loop heating $h_{Poyn}$ that would be produced if all the 
magnetic field energy input corresponding to the Poynting flux 
$S_{phot}$ was converted to thermal energy. This quantity is 
computed for comparison with the heating terms derived using particle acceleration and presented in Section~6. As $h_{Poynt}$ is an average value 
over time and space, it does not simulate the intermittent nature of nanoflares. It is given as
\begin{equation}\label{heat_poynt}
h_{Poyn} =\frac{A_{phot}\, (|S_{foot1}\, +\, S_{foot2}\, |)}{V_{loop}}
\end{equation}
In Equation~(\ref{heat_poynt}) $A_{phot}$ is the cross-section at the 
footpoints of each loop, which is equal to the square of the 
magnetogram's binned pixel size, i.e. 0.6~Mm$^2$. $V_{loop}$ is the 
loop volume (Figure~\ref{fig:Bmean}c). Here, $V_{loop}$ is calculated by integrating
the volume of infinitesimal cylinders corresponding to the different cross-sections along the loop.
We note that in subsequent calculations the cross-section is allowed to vary {\it along} 
each loop in order to conserve the magnetic flux passing through it. The constant magnetic flux of each loop
equals the average magnetic flux calculated at its two footpoints.
In Figure~\ref{fig:Bmean}c, we performed a logarithmic fit which shows that the volume as a function
of the loop length is described by $V_{loop} = 0.07\, L^{1.85}$. This 
expression is used in Section~6 to describe a loop heating function.

In Figure~\ref{fig:Bmean}d we see that $h_{Poyn}$ is in the range of 
$10^{-5}$ to 0.4~$\mathrm{erg\, s^{-1}\, cm^{-3}}$. Moreover, $h_{Poyn}$ is decreasing 
as a function of loop length $L$ due to the latter's inverse dependence on 
the loop volume $V_{loop}$. In the following, $h_{Poyn}$ will be compared 
with the heating rates computed by the particles' acceleration.

We finally note that, instead of the sum $|S_{foot1}+S_{foot2}|$, 
we have also made calculations using $|S_{foot1}|+|S_{foot2}|$ as an upper estimate of the 
total Poynting flux $S_{phot}$ through a loop. Both expressions lead
to quite similar results. Hereafter we will only refer to calculations 
performed by the choice $|S_{foot1}+S_{foot2}|$.

\section{Current sheet modeling}

\begin{figure}
\epsscale{.80}
\plotone{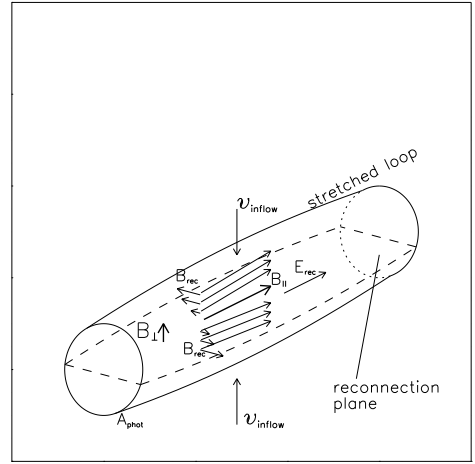}
\caption{Stretched cylindroidal coronal loop. The lateral areas
$A_{phot}$ correspond to the loop footpoints. The current sheet
is formed at a reconnection plane  parallel to the cylindroid's
main axis. The magnetic field changes orientation in the
half-cylindroids defined by the reconnection plane. The magnetic
field projection perpendicular to the cylinder's axis forms the
reconnecting component $B_{rec}$ while the projection parallel to the
axis forms the parallel magnetic field component
$B_\parallel$. The component $B_\perp$ is perpendicular to the current sheet plane.}
\label{fig:rec_draw}
\end{figure}

We now proceed to model current sheets formed along our extrapolated coronal 
loops. While a reasonable assumption is that each loop should 
include several current sheets at sub-resolution scales, it 
will be shown below that the resulting coronal heating by 
the cumulative result of multiple current sheets distributed 
inside the loop is equivalent to the heating from a 
single \lq average\rq\ current sheet extending from footpoint 
to footpoint and consisting of a tangential discontinuity with
a variable magnetic field vector across its surface.
Figure~\ref{fig:rec_draw} presents a schematic view of a stretched 
out cylindroidal coronal loop that contains one 
such current sheet. Since the magnetic flux is conserved along 
each loop, the loop's cross-section increases as we move from 
one of the footpoints towards the loop's apex.
The current sheet is on the reconnection plane (dashed cut) which 
includes the cylindroid's axis. In Figure~\ref{fig:rec_draw}, the 
reconnection plane extends through the entire cylindroid. As shown in Section~5, this does not 
influence the resulting heating. The plane 
of the current sheet coincides with the plane of the discontinuity 
formed due to shearing motions at the photospheric level which 
presumably lead to the formation of two distinct magnetic-flux
domains within the loop volume.

The magnetic field vectors in Figure~\ref{fig:rec_draw} are 
depicted with different orientations above and below the reconnection 
plane. The angle $\theta_D$ formed between two adjacent flux tubes at the 
discontinuity is assumed to be equal to twice the Parker angle. 

These magnetic flux tubes, which form individual coronal loops, have a linear cross section smaller than 100 to 200~km
\citep{Cirt13,Chen13}. Our magnetogram data, binned to 1.1\arcsec, cannot resolve these sub-resolution structures  i.e. we cannot have 
any observational evidence on the value of the angle $\theta_D$. Given this limitation,
in the present study we assigned a value of the angle $\theta_D$  to each loop, picked from 
a uniform, distribution between $8\degr$ and $50\degr$. The minimum value $8\degr$ 
corresponds to the upper limit of the mean inclination of magnetic fields in simulations of MHD-turbulence \citep{Rap07} while the upper value
includes (twice the) Parker angles of $40\degr$ and $45\degr$ derived from numerical simulations \citep{Dah09,Gal96}.
Moreover, it includes the typical value of $40\degr$ found from energy considerations \citep{Klim06}.
We also attempted to calculate $\theta_D$ using the scaling law derived in \citet{Rap07}. The results of the various
calculations will be discussed in detail in section~6.

The shearing and twisting motions at the photospheric level are assumed such 
that the Poynting flux $S_{phot}$ has a direction of injection inwards, i.e., into the sheet. For 
simplicity, we also assume that the current sheet is described by a Harris 
type geometry. In this geometry, two out of the three sheet's 
magnetic field components can be derived from the loop mean magnetic field $\bar B$, and $\theta_D$: 
i) the magnetic field component perpendicular to the loop axis and parallel to the current sheet plane is assumed 
to increase linearly with distance from the current sheet surface. Its 
maximum value $B_{rec}$, at the edges of the current sheet 
(see Figure~\ref{fig:rec_draw}), is given by $B_{rec}=\bar B \sin(\theta_D/2)$. 
ii) The magnetic field component parallel to the loop axis 
$B_{||}$ corresponds to the current sheet's guide field component. This component 
is assumed constant at each point inside the current sheet, and it is
expressed as $B_\parallel=\bar B \cos(\theta_D/2)$. The dimensionless 
parameter $\xi_\parallel = B_\parallel/B_{rec}$, is important 
in current sheet literature \citep{Efth05,Litv00}. 
In the present modeling it is given by $\xi_\parallel= cot(\theta_D/2)$. 
According to the above definitions, for $\theta_D=0$, the tangential discontinuity vanishes as $B_{rec}=0$
and $B_\parallel=\bar B$. For $\theta_D=\pi$, we end up to an anti-parallel reconnection with $B_\parallel=0$ and $B_{rec}=\bar B$. 
For the adopted range of $\theta_D$ between $8\degr$ and $50\degr$ we end-up with $\xi_\parallel$ between 14.3 and 2.15 respectively.
With the above settings, the injected Poynting flux $S_{rec}$ is 
expressed by the energy conservation equation as
\begin{equation}\label{S_rec_phot}
2\,S_{rec}A_{rec}\, =\, A_{phot}\, (|S_{foot1}\, + \, S_{foot2}|)
\end{equation}

In Equation~(\ref{S_rec_phot}), $A_{rec}$ is the current sheet area and the 
factor 2 accounts for the fact that the Poynting flux is injected from both sides 
of the current sheet. Therefore, in Figure~\ref{fig:rec_draw}, $A_{rec}$ 
corresponds to the area of the dashed current sheet. 
$A_{rec}$ is calculated along each loop by integration, taking into 
account that the loop diameter normal to the loop's axis varies 
along the loop. $A_{rec}$ takes values in the range from 10 to 
$760$~Mm$^2$, with longer loops exhibiting larger values of 
$A_{rec}$.
\begin{figure}
\epsscale{.80}
\plotone{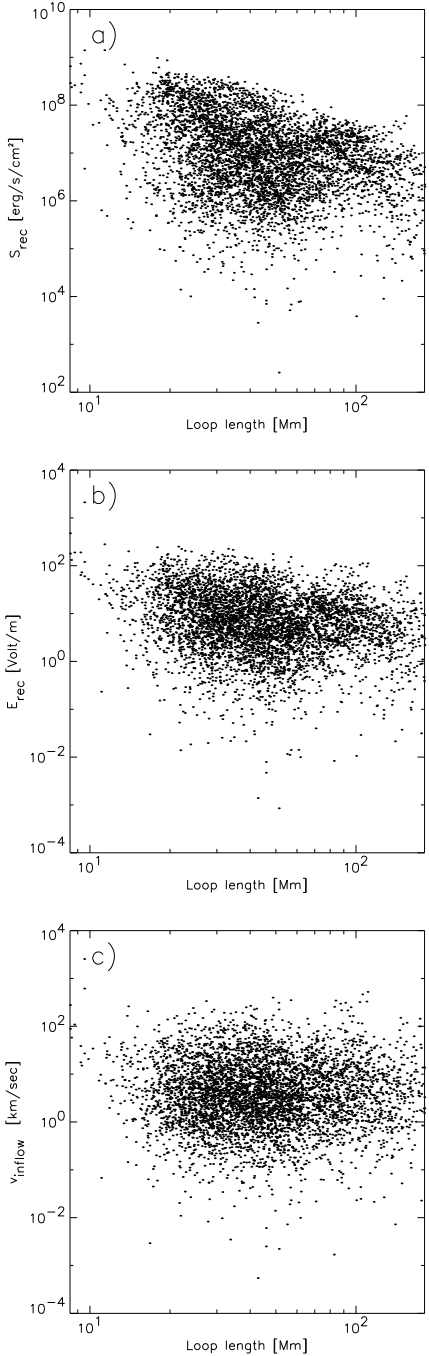}
\caption{Scatter plots of (a) the Poynting flux injected (b) 
the reconnecting electric field, and (c) the plasma velocity 
inflow inside each of our considered current sheets.}
\label{fig:Poynt}
\end{figure}

The inflowing Poynting flux is directly related to the induced electric 
field $\vec E_{rec}$ which accelerates particles inside the current 
sheet. The electric field is given by
\begin{equation}\label{E_rec}
\vec E_{rec}\,=\, -  \frac{1}{c}  ( \vec v_{inflow} \times \vec B_{rec})
\end{equation}
where $\vec v_{inflow}$ is the velocity of the plasma injected in the 
current sheet. The velocity $\vec v_{inflow}$ can be derived from the expression of the injected 
Poynting flux $S_{rec}$ via
\begin{equation}\label{S_rec}
\vec S_{rec}\,=\,  \frac{c}{4\, \pi}  (\vec v_{inflow} 
\times \vec B_{rec}) \times \vec B_{rec}
\end{equation}

Figure~\ref{fig:Poynt} presents the results of the above calculations. 
Figure~\ref{fig:Poynt}a shows that the injected $S_{rec}$ values are 
in the range $10^4$ to $2 \times\, 10^9$~$\mathrm{erg\, s^{-1}\,cm^{-2}}$. Furthermore, $S_{rec}$ 
is decreasing with loop length because it is inversely 
proportional to $A_{rec}$. For short loops ($L < 30$~Mm), the mean injected 
Poynting flux is $S_{rec}\simeq 7.5\times\ 10^7$~$\mathrm{erg\, s^{-1}\,cm^{-2}}$, for intermediate size loops ($30$~Mm$<L<100$~Mm)
we find $S_{rec}\simeq\, 2.2\times\, 10^7$~$\mathrm{erg\, s^{-1}\,cm^{-2}}$, while for 
long loops ($L>100$~Mm) $S_{rec}\simeq 7\times\, 10^6$~$\mathrm{erg\, s^{-1}\,cm^{-2}}$. 
Here we assume that, in an electron-proton plasma at thermal equilibrium,  
electrons will be accelerated practically without friction force 
(due to Coulomb interactions with the protons), because the electric field 
applied to the plasma is much larger than the Dreicer electric field 
computed for the standard coronal temperatures and plasma densities \citep{Dr59,Ben93}.
In general, it is expected that electric fields appearing
during solar-flare reconnection events are much larger than the Dreicer 
electric field \citep{Mar90}. An explicit computation of the Dreicer 
field values of our model is given at the end of this section.

As seen in Figure~\ref{fig:Poynt}b, 99\% of the electric field values are between 0.01 and 100~V/m.
The electric fields are also decreasing 
for increasing loop length. Moreover, 96\% of the  inflow velocity values (see 
Figure~\ref{fig:Poynt}c) are in the range 0.1 to 100~\kms\, but 
there appears to be no correlation between $v_{inflow}$ and the loop length.

The next important parameter to compute is the magnetic 
field component $B_{\perp}$ which is perpendicular to the plane of the current sheet (see Figure~\ref{fig:rec_draw}).
Assuming consistency between the current derived from Amp\'ere's law 
and the electric current produced by the accelerated protons, considering that they carry
most of the particle energy, one finds \citep{Eas72,Mar90,Litv96} :
\begin{equation}\label{bperp}
v_A B_\perp\,=\, \sqrt{2} v_{inflow} B_{rec}
\end{equation}
where $v_A$ is the Alfv\'en speed. The above equation is valid for 
$\xi_\parallel =0$. In the absence of a guide component, the reconnecting 
component $B_{rec}$ equals the average magnetic field $\bar B$ so that, in Equation~(\ref{bperp}) the Alfv\'en
speed can be expressed as $v_A= B_{rec}/\sqrt{4 \pi m_p n_e}$.
Equation~(\ref{bperp}) differs from the one found in 
\cite{Eas72}, \citet{Mar90}, or \citet{Litv96} by a factor $\sqrt{2}$.
This numerical factor is introduced because we estimate the proton kinetic energy 
twice as much as in the above mentioned works, for reasons explained in Section~5.
For $\xi_\parallel \neq 0$, on the other hand, 
the reconnecting field component is not equal to the mean magnetic 
field. Therefore, $\bar B$ should replace $B_{rec}$ in the Alfv\'en speed expression 
through the definition $B_{rec}=\bar B \sin(\theta_D/2)$. The term $\sin(\theta_D/2)$ appears in 
Equation~(\ref{bperp}) which becomes:
\begin{equation}\label{bperp2}
v_A B_\perp\,=\, v_{inflow} B_{rec} \frac{\sqrt{2}}{\sin \frac{\theta_D}{2}}
\end{equation}
Let us note that in the limit  $\theta_D=\pi$, (anti-parallel reconnection), Equation~(\ref{bperp2}) becomes identical to Equation~(\ref{bperp}).

In order to compute the Alfv\'en speed  we use the \citet{Ros78} scaling-laws to determine the electron density for each loop assuming a maximum temperature
of $T_{max}=10^6$~K. These scaling laws are valid in a strict sence for hydrostatic atmospheres. Their use in the present context will be commented in Sect.~9. Because of the relatively low 
value of $T_{max}$ we have also low electron densities, i.e., in the range $\mathrm{2\times 10^8~cm^{-3}}$ up to $\mathrm{6\times 10^9~cm^{-3}}$.
\begin{figure}
\epsscale{.7}
\plotone{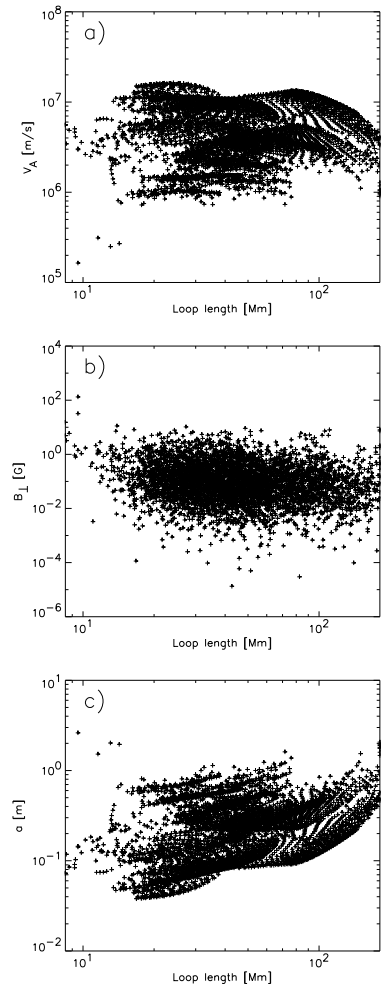}
\caption{Scatter plots of (a) the Alfv\'en speed, (b) the 
perpendicular component $B_{\perp}$, and (c) the current sheet 
thickness, as a function of the loop length, for all current sheets considered.}
\label{fig:Alfven}
\end{figure}

Figure~\ref{fig:Alfven}a shows the Alfv\'en speed for all our current sheets. 
The values of $v_A$ are in the range 
between $10^3$ and $1.6 \times 10^4$~\kms which are expected values for the coronal plasma. 
From Equation~(\ref{bperp2}) we compute $B_\perp$ for 
all the current sheets (Figure~\ref{fig:Alfven}b), with 80\% of the values in the 
range of 0.01  to $1$~Gauss. The average $B_\perp$
is 0.3~Gauss and its standard deviation 0.85~Gauss. Moreover, for the 
dimensionless quantity $\xi_\perp = B_\perp/ B_{rec}$, 98\% of the values are in the range of $10^{-5}$ to 1.

Finally, we estimate the current sheet thickness as follows: according 
to \citet{Cas08}, a current sheet becomes collisionless and susceptible 
to reconnection when its thickness $a$ becomes of the order of the 
ion gyro-radius. However, when the guide magnetic field component 
is non-zero, the instability leading to reconnection occurs
when the current sheet thickness becomes of the order of the Hall 
scale $a\,=\, v_{s}(T)/\omega_{ci}$, \citep{Cas07}, where $v_s$ is the 
sound speed calculated for a  plasma temperature $T=10^6$~K and $\omega_{ci}$ is the 
ion-cyclotron frequency. The current sheet thickness can be also 
expressed as $a\,=\,5.69\times 10^{-8} v_{s} \bar B^{-1} \frac{m_p}{m_e}$.
The resulting thickness turns to be of the order of the electron 
gyroradius. Figure \ref{fig:Alfven}c shows that the computed 
thicknesses in our current sheet sample range from 0.01 to 1~m, 
with the longer loops supporting thicker current sheets. Using the 
electron density $n_e$ and temperature $T=10^6$~K we also calculated 
the Dreicer electric field, which is given by the expression $E_D\,=\, 6.06\times\, 10^{-6} n_e/T$~Volt/m \citep{Ben93}. 
Our computed electric fields $E_{rec}$, are larger 
than the Dreicer electric fields by factors ranging from 1 up to 
$10^4$ for 97\% of the cases. We conclude that 
the assumption of collisionless acceleration of the particles in 
our modeled current sheets is consistent with the adopted plasma 
parameters.

\section{Particle acceleration}
In this section we compute estimates of the kinetic energy gain 
of electrons and protons accelerated through the loop current 
sheets considered in the previous section. 
In our approach, charged particles enter the current sheet with a velocity $v_{inflow}$ and are 
accelerated by the induced DC-electric field $E_{rec}$. Inside the current sheet, particles follow a trajectory 
which depends on the initial velocity and on the strength of the electric and magnetic fields. Particles are ejected 
before they can travel along the total length of the current sheet, which in our case is equal to the loop length, due to the Lorenz force raised by the $B_\perp$ component \citep{Spei65}. 
For current sheets with $\xi_\parallel > 1$, particles will follow adiabatic orbits with a very small amount of chaos \citep{Efth05}. 
In the present study we did not examine the possibility that a particle interacts with more than one current sheet. Therefore, after the ejection from the current sheet, 
particles move along the magnetic field lines without any further acceleration. 
The particles' kinetic energy gain is proportional to the electric field strength multiplied by the final acceleration length 
(along the electric field, see Fig.~\ref{fig:rec_draw}). As the  electric fields $E_{rec}$ are super-Dreicer, 
collisions are ignored. Therefore, to estimate the final kinetic energy one needs to compute the 
orbit as long as the particle is under the influence of the current sheet electric and magnetic fields.
\citet{Efth05} derived
an analytical expression for the kinetic energy range $E_k$
of accelerated particles inside 
a Harris type current sheet as a function of the initial particle 
energy $E_{k0}$ and of the current sheet's parameters (field strengths and 
thickness). This expression reads
\begin{align}\label{kinetic}
E_{k j}\,=\, E_{k0}\,+  \frac{E_{rec} }{B_\perp^2} 
\Big( e\, B_\parallel B_\perp a + m_j E_{rec}  \nonumber \\
\pm  \sqrt{2 e\, m_j\, a\, B_\parallel B_\perp  E_{rec} 
+ m_j^2 E_{rec}^2 + 2\, m_j\, B_\perp^2  E_{k0} } \Big)
\end{align}
In Equation~(\ref{kinetic}), $j$ represents the particle species 
(electrons or protons). The two values for the $\pm$ sign 
define the energy range around a mean kinetic energy. 
In the sequel we consider only the upper limit of the particles' kinetic 
energy range (plus sign in Equation~\ref{kinetic}). Furthermore we consider that the 
initial particle energies obey a Maxwellian distribution at a temperature of 10$^6$~K. 

\begin{figure}
\epsscale{.7}
\plotone{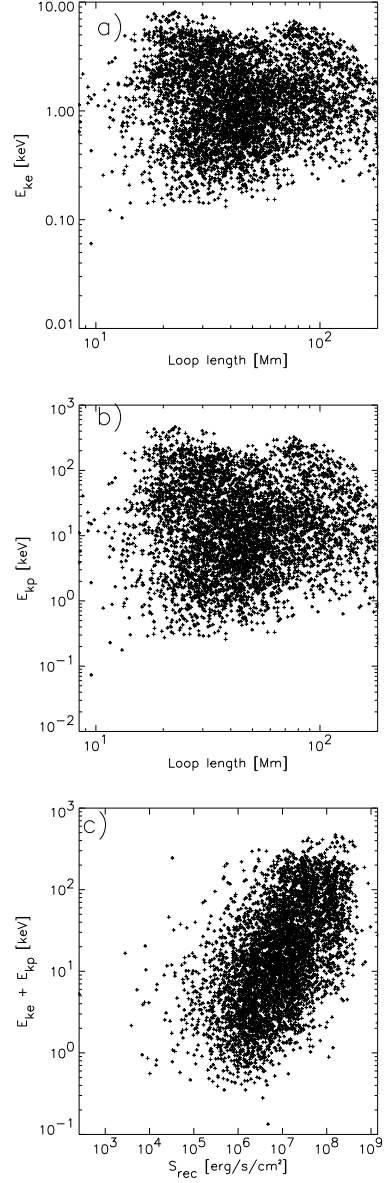}
\caption{Scatter plots of the final kinetic energy for electrons (panel a), and protons (panel b). Panel c) shows the total kinetic energy of electrons and protons as a function of the Poynting flux entering the current sheet.
}\label{fig:kinet}
\end{figure}

In \citet{Efth05}, the analytical expression is more cumbersome than in Eq.~\ref{kinetic} as it includes explicitly $I_2$, an integral of the motion of particle orbits, resulting from the translational symmetry of the Harris type sheet geometry along $\vec E_{rec}$ (see Figure~\ref{fig:rec_draw} and \citet{Efth05}, \citet{Litv93}). In Equation~\ref{kinetic}, we assume, for simplicity, that $I_2=0$. However, even with this restriction, Equation~(\ref{kinetic}) gives a good estimate of the final kinetic energy of particles. 
The first two terms in Equation~(\ref{kinetic}),
\begin{equation}\label{kinetic_simple}
E_{ke}=E_{k0}+\frac{e E_{rec} B_\parallel a}{B_\perp}
\end{equation}
are sufficient to describe the electrons average final kinetic energy \citep{Litv00} while for the protons, the kinetic energy is well described by the expression \citep{Litv00,Efth05}
\begin{align}\label{kinetic_protons}
E_{kp}\,=\,2 m_p c^2\, \Big( \frac{E_{rec}}{B_\perp} \Big)^2
\end{align}
where the factor~2 originates because in Equation~(\ref{kinetic}) the $m_p E_{rec}$ term is found twice and causes the $\sqrt{2}$ factor in Equations~\ref{bperp},\ref{bperp2}. The initial particle kinetic energy, corresponding to a Maxwellian kinetic energy for the selected temperature of the order of 0.04~keV, is, on the average, 40 and 800 times smaller than the final kinetic energy of the electrons and protons respectively, after the acceleration process. The final electron kinetic energies are of the order 0.1 to 8~keV while final proton kinetic energies are in the range 0.3~keV to 470~keV (Figure~\ref{fig:kinet}).

Figure~\ref{fig:kinet}a, shows the kinetic energy distribution of electrons while Figure~\ref{fig:kinet}b shows the kinetic energy distribution of protons. The largest kinetic energies are found for loops with lengths ranging between 15 and 30~Mm. In Figures~\ref{fig:kinet}a,b the kinetic energy scatter plots for electrons and protons have a similar shape. This is because the kinetic energy of electrons depends on the Alfv\'en velocity while the kinetic energy of protons depends on the square of the Alfv\'en velocity. This dependence on the Alfv\'en velocity, which is implicit in Eqs~\ref{kinetic_simple},\ref{kinetic_protons} comes from the definition of $B_\perp$  in Eq.~\ref{bperp}.
This dependence becomes explicit after the following calculation. Combining Equations~(\ref{E_rec}),(\ref{S_rec}),(\ref{bperp2}),(\ref{kinetic_simple}), (\ref{kinetic_protons}) as well as the definitions of the thickness $a$ and the $B_\parallel$ component, we obtain the dependence of the sum of electron and proton kinetic energy gain on the initial parameters as :

\begin{align}\label{praxeis}
E_k=\frac{e\, E_{rec}B_\parallel a}{B_\perp}\,+\, 2\, m_p c^2\, \Big(\frac{E_{rec}}{B_\perp}\Big)^2 =\nonumber \\
=\frac{e}{c\,\sqrt{2}}\, \bar B \cos \frac{\theta_D}{2}\, a v_A\, \sin\frac{ \theta_D}{2}\, +\, 2 m_p c^2\, \Big( \frac{ v_A\, \sin\frac{ \theta_D}{2}}{c\sqrt{2}}\Big)^2 \nonumber \\
\Rightarrow  E_k = \frac{e}{c\,2\sqrt{2}} c_1 v_s(T) v_A\, \sin\,\theta_D \,+\, m_p  v_A^2\, \sin^2\frac{\theta_D}{2}
\end{align}
Equation~\ref{praxeis} holds for $0<\theta_D<\pi$ since for $\theta_D=0$ we have no current sheet while for $\theta_D=\pi$ the reconnection is anti-parallel, ($\xi_\parallel=0$) and one should use Eq.~\ref{kinetic}. Moreover we 
omitted the initial kinetic energy $E_{k0}$ from Eq.~\ref{praxeis}. In Equation~\ref{praxeis}, the constant $c_1=5.8 \times 10^{-8} \frac{m_p}{m_e}$ originates from the expression for the thickness $a$ \citep{Cas07}, introduced in Section~4.
Equation~(\ref{praxeis}) shows that the final kinetic energy $E_k$ of the accelerated particles depends on the Alfv\'en velocity, the plasma density and the discontinuity angle $\theta_D$. The sound velocity $v_s(T)$ is introduced
through the thickness $a$ of the current sheet. The kinetic energy $E_k$ is implicitly related to $S_{rec}$ as seen in Figure~\ref{fig:kinet}c, where we observe that $S_{rec}$ and $E_k$ are well correlated.
Moreover, the electron kinetic energy as a function of the mean magnetic field at the loop footpoints is well fitted by power law
$E_{ke} \propto \bar B_{foot}^{0.65}$. This means that for a different active region, with stronger photospheric magnetic fields, reaching, for example
5000~Gauss, we expect that electron kinetic energies can reach up to 10~keV.  Another important aspect is that $E_k$ is independent from the surface area $A_{rec}$ of the current sheet. This means that the initial selection of the current sheet so as to cover the entire surface of the loop cross-section does not influence the resulting kinetic energies. This allowed us to choose, for each loop, a single current sheet with a simple geometry to describe particles' acceleration.

In addition, the acceleration length $z_{max}$ of the particles is an important parameter in our model. The acceleration length is the distance along the electric field covered by the particles and it is defined via the expression 
$E_k = e\, E_{rec}\, z_{max}$ for both particles species. $z_{max}$ must obviously be smaller than the current sheet length or equivalently the coronal loop length. We found that for electrons, the ratio $z_{max}/L$ is between $10^{-7}$ and $10^{-3}$ in 97\%  of the cases, while for protons it is between $10^{-6}$  and 0.01 in 90\% of the cases. For protons, the average value of $z_{max}/L$ is $10^{-3}$ with a standard deviation of 0.1. A short acceleration length relative to the loop length reduces the electric current intensity which, in turn, reduces the induced magnetic fields generated by the electric current of the accelerated particles \citep{Mar90,Litv96}.

We performed test particle simulations for the acceleration of electrons and protons similar to the ones presented in \cite{Gont07} and \cite{Ana08}. In these simulations, single particles, having initially a random $10^6$~K thermal velocity, enter a Harris type current sheet with given initial parameters $(a,B_\parallel,B_\perp,E_{rec},B_{rec})$. Solving the equations of motion, a particle's orbit is traced until the particle leaves the current sheet, at half-width distance from the inversion surface. 
Running simulations for all 5000 current sheets used in this study is unrealistic, as they are time consuming.
Therefore we selected 10 representative values of electric fields $E_{rec}$, reconnecting, parallel and perpendicular magnetic components ($B_0, B_\perp, B_\parallel$) and current sheet thicknesses $a$ from our calculated distributions.
The resulting kinetic energies are in agreement with Equation~(\ref{kinetic}). All 1000 electrons used in each simulation are accelerated to kinetic energies of the order inferred by Equation~(\ref{kinetic_simple}). However, most protons cross the current sheet with no energy gain and only roughly 10\% of protons are accelerated. The accelerated protons are the ones having an initial velocity with a particular orientation relative to the current sheet. The fact that only a fraction of protons are accelerated was also found in \cite{Gont07}.

To recapitulate, to estimate the particles' kinetic energies, we use observations of photospheric magnetic fields and inferences of the photospheric horizontal velocities. From them, we calculate the photospheric Poynting flux $S_{phot}$ (Equation~(\ref{Poynt_phot_eq})). From a potential magnetic field extrapolation, we select 5000 closed loops, on each of which we calculate the mean magnetic field $\bar B$.
From this, and by virtue of the assumptions described before, we calculate the parameters $B_{rec}$, $B_\parallel$, $v_{inflow}$, $E_{rec}$, $a$ and $B_\perp$ necessary and sufficient to calculate kinetic energies of accelerated particles.

\section{Loop heating due to accelerated particles}
We compute the heating of coronal loops assuming that it is produced solely by the ensuing thermalization of the accelerated particles.  We assume that both electrons and protons participate in this process, since  
both particles' energy is released once they hit the dense chromospheric layer. Protons with energies ranging between 100~keV and 5~MeV can produce chromospheric evaporation thus participating in coronal heating \citep{Emslie96}.
In our simulation, only 10\% of the protons' kinetic energies are above 100~keV. Nevertheless we assume that even protons with kinetic energy less than 100~keV participate in the loop heating.

The thermalization of the accelerated particles is a complex process which we do not model in detail. Instead, we use the following phenomenological expression for the heating rate :
\begin{equation}\label{heating}
Q\,=\, \frac{(E_{ke}+f\, E_{kp}) n_e}{t_{rec} } \frac{V_{rec}}{V_{loop}}
\end{equation}
In Equation~(\ref{heating}), the heating rate $Q$ of a loop with a pre-nanoflare electron density $n_e$, corresponds to the thermalization of current-sheet accelerated electrons and protons to kinetic energies $E_{ke}$ and $E_{kp}$ respectively. A neutral, fully ionized plasma $(n_e\,=\,n_p)$ is assumed.
Here $V_{rec}=2\, v_{inflow}\, t_{rec}\, A_{rec}$ is the plasma volume injected inside the current sheets of total surface $A_{rec}$ during the reconnection time $t_{rec}$. As reconnection time we define the duration of the reconnection event. The fraction  $V_{rec}/V_{loop}$ indicates that the heating produced inside the current sheets is redistributed to the loop volume $V_{loop}$. The efficiency factor $f=0.1$ means that only 10\% of protons are accelerated, as found by the test particle simulations presented in the end of section~5.

In subsequent calculations we use Equations~\ref{S_rec_phot}, \ref{E_rec}, \ref{praxeis} and assume that the loop volume is $V_{loop} = C_4 L^{-p}$, where $C_4=50.1$, $L$ the loop length, and $p=1.85$ as found by the fit to the calculated loop volumes in Figure~\ref{fig:Bmean}c. The heating $Q$ is expressed (in $\mathrm{erg\, s^{-1} cm^{-3}}$) as :
\begin{align}\label{heat_praxeis}
Q\,=\, \frac{4 \pi}{c} \Big(\frac{e}{c\,2\sqrt{2}}\, c_1\, \sin\theta_D\, v_s(T) v_A\,+ \nonumber \\
\, f m_p v_A^2\, \sin^2\frac{\theta_D}{2} \Big) \frac{n_e \bar S_{foot} A_{phot}}{B_{rec}^2 C_4 L^p}
\end{align}
Here $\bar S_{foot} A_{phot}$ is the average photospheric Poynting flux times the footpoint cross-section. In Equation~(\ref{heat_praxeis}) we can see that the heating rate $Q$ does not depend on $t_{rec}$ and that the electron and proton terms have a different form of dependence on $v_A$.
We can also express $Q$ as a function of $L$ if we replace the electron density $n_e$ by the expression given in \citet{Ros78}.
The individual heating rates due to electrons and protons have a power law dependence on $L$ but with a slightly different exponent
\begin{align}\label{heat_praxeis2}
Q_e \propto \bar S_{foot} \frac{ \cot\frac{\theta_D}{2} T v_s(T)}{\bar B  L^{\frac{1+2p}{2}}}  \\
Q_p \propto \bar S_{foot} \frac{1} {L^p}
\end{align}
\begin{figure}[!h]
\epsscale{.8}
\plotone{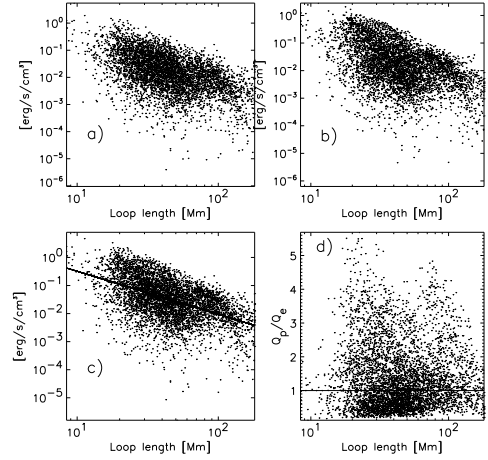}
\caption{Heating rates produced by electrons (panel a), protons (panel b), and by both particle types (panel c) as a function of loop length. Panel d shows the ratio of proton heating over electron heating as a function of loop length. Panel c) also shows a linear fit to the log of the heating rate distribution.}\label{heat_functions:fig}
\end{figure}
Figure~\ref{heat_functions:fig} shows the heating rates due to electrons (Figure~\ref{heat_functions:fig}a), protons (Figure~\ref{heat_functions:fig}b) and both types of particles (Figure~\ref{heat_functions:fig}c) 
as a function of loop length. 
Despite the larger scatter, the similar power law dependence of the heating produced by each type of particles, shown in Equation~(\ref{heat_praxeis2}), is also visible in the scatter plots of Figure~\ref{heat_functions:fig}a-c.
The heating-flux ratio $Q_p/Q_e$ seen in Figure~\ref{heat_functions:fig}d takes values in the range 0.1 to 5 with 50\% of the values larger than 1. The heating rate function from the two particle species (Figure~\ref{heat_functions:fig}c) is in the range of $10^{-4}$ to 1~$\mathrm{erg\, s^{-1} cm^{-3}}$. 
We also show a linear fit to the logarithm of the values in the scatter plot. The calculated power law index is of $c_2=-1.5$. This index differs from $p=1.85$, the exponent of $L$ in Equation~(\ref{heat_praxeis2}). The reason is that $\bar S_{foot}$, which appears in Equation~(\ref{heat_praxeis2}) has also a weak dependence on $L$, with a positive power law index $a=0.35$, which influences the resulting fit. In \citet{Man00}, the heating rate deduced from loops observed in X-rays with the SXT telescope on Yohkoh has a power law relation with the loop length, with exponents in the interval [-4.5,-1] and most probable value $\simeq -2$. Therefore, the exponent value of $-1.5$ presently derived is fairly consistent with observations. 
\begin{figure}[!h]
\plotone{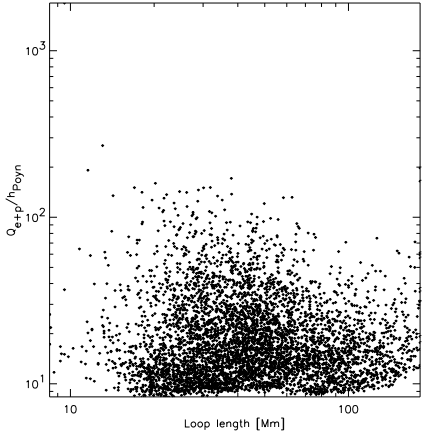}
\epsscale{.8}
\caption{Ratio of the heating rates due to particles over $h_{Poynt}$, the energy rate corresponding to the Poynting flux.}\label{heat_ratio:fig}
\end{figure}
In Figure~\ref{heat_ratio:fig}, we plot the ratio of the total heating $Q_{p+e}=Q_p\,+\,Q_e$, due to electrons and protons, over the value of the Poynting flux function, $h_{Poynt}$ injected inside the loops, (plotted in Figure~\ref{fig:Bmean}). 
We observe that 95\% of the values $Q_{p+e}/h_{Poynt}$, , are in the range of 9 to 50 with an average of 21. 

Thus, the heating rate found in our model by consideration of particle acceleration overestimates the one induced by the value of the Poynting flux by an average factor $\simeq 20$. 
This appears at first as a large inconsistency. However, as pointed out in \citet{Ros10} or \citet{Birn07} (p. 287) this inconsistency should be regarded as a consequence of the inherent lack of self-consistency in all methods estimating particle acceleration via the trajectories of test particles. In fact, in MHD simulations of coronal heating \citep{Hen96, Gal96, Rap07} one finds $Q_{p+e}/h_{Poynt}\simeq\, 1$. However, in such  simulations the balance is restored partly due to an additional feedback mechanism caused by the on-going twisting and relaxation of magnetic fields. On the other hand
the effects of magnetic reconnection on the acceleration of particles cannot be captured by such simulations. At any rate, our present results show that the thermalization of the loops is indeed possible, at least as shown by order of magnitude estimates, via the conversion of magnetic energy to particles' kinetic energy during magnetic reconnection.
\begin{figure}[!h]
\epsscale{1.0}
\plotone{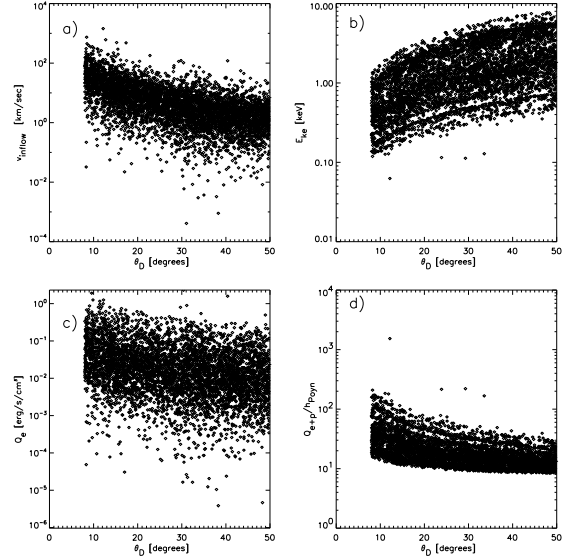}
\caption{Some key quantities of our model as a function of $\theta_D$. Panel a) shows the $v_{inflow}$ velocity, panel b) shows the final kinetic energy of electrons, panel c) shows the
heating due to the accelerated electrons and panel d) the ratio of heating due to accelerated electrons and protons over the heating corresponding to the supplied Poynting flux.
In all panels, the horizontal axis shows $\theta_D$ values for each loop.}\label{angle_dependence:fig}
\end{figure}

\begin{figure}
\epsscale{1.0}
\plotone{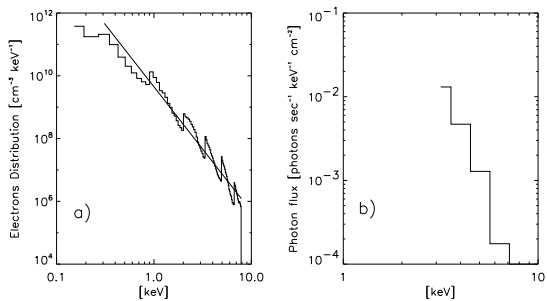}
\caption{Average kinetic energy distribution from the 5000 selected loops (panel a) and the corresponding X-ray spectrum (panel b) calculated according to the thick target approximation.}\label{x-ray_flux:fig}
\end{figure}
One more feature to notice is that a possible lifetime of a single current sheet could be of the order of the Alfv\'en crossing time, which is equal to the fraction of the current sheet length over the Alfv\'en speed \citep{Klim06}. In our model this time ranges from 1~s to 135~s. On the other hand, hydrodynamic simulations of nanoflares consider storm of many nanoflares with a duration of 50~s to 500~s \citep{Klim06}. Let us note that the ratio $V_{rec}/V_{loop}$ is less than 1 for 95\% of our current 
sheets, if we set their lifetimes equal to the Alfv\'en crossing time. This means that for a 5\% of the loops, reconnection would end practically because of the lack of sufficient plasma inflow.
For the other cases, only a fraction of the total number of particles stored in each loop was able to cross the current sheet during the reconnection event. Moreover, the time needed for an individual particle to cross the current sheet is much smaller than any of the above time scales. For electrons, the particle acceleration time is in the range $10^{-6}$~s to 0.01~s while for protons it is in the range $10^{-4}$~s to 1~s. In 98\% of cases the acceleration times of protons turn to be smaller than one tenth of the corresponding current sheet lifetime. This justifies our assumption that the particles interact with a quasi-stationary environment during the acceleration phase. 

An additional remark is that our computed accelerated particles' kinetic energies and heating functions are still valid if, instead of a monolithic current sheet per loop, we assumed a large number of current sheets, as long as these sheets are formed by the same discontinuity angle $\theta_D$, the same magnetic field components $B_{rec}, B_\perp, B_\parallel$, the same thickness and they are longer than the corresponding acceleration length $z_{max}$. Moreover, we only consider particles interacting with only one current sheet during their travel inside the loop. Thus Equation~(\ref{heating}) describes the heating per current sheet which  is proportional to a fraction $S_{foot} A_{phot}$ of the heating rate supplied. In
Section~\ref{section_hydro} we will calculate the hydrodynamic response of the studied loops under the effect of a number of current sheets producing the same collective heating effect with the one described by Equation~(\ref{heating}).

To summarize the results of Sections~5 and 6, Figure~\ref{angle_dependence:fig}, shows some model parameters as a function of $\theta_D$. These parameters were presented in previous figures as a function of the loop length. The inflow velocity $v_{infow}$,  (Fig.~\ref{angle_dependence:fig}a) has higher values for a lower $\theta_D$. In Fig.~\ref{angle_dependence:fig}b, the kinetic energies of electrons decrease with increasing $\theta_D$ according to Eq.~\ref{praxeis}, where 
the first term in the last sum depends on $\sin\,\theta_D$.
On the other hand, $Q_e$ is higher for lower $\theta_D$ (Fig.~\ref{angle_dependence:fig}c), as expected from Eq.~\ref{heat_praxeis2}. Finally, in Fig.~\ref{angle_dependence:fig}d one can see that for higher $\theta_D$,  $Q_{e+p}/h_{Poynt}$ tends asymptotically to a lower limit $\simeq 8$.

We also calculated $\theta_D$ according to a scaling law given in Eq.~10 in \citet{Rap07}. The main parameter in this equation is the ratio of photospheric velocity $v_{ph}$ over the alfvenic velocity $v_A$. For our data, the derived $\theta_D$ were in all cases less than 4$\degr$. The small $\theta_D$ values are due to the low ratio of photospheric over alfvenic velocities in our model. Alfv\'en velocities are high due to the low $n_e$  calculated using the scaling law of \citet{Ros78} with a relatively low temperature of $10^6$~K. As shown in Fig.~\ref{angle_dependence:fig}, such small $\theta_D$ values correspond to even larger $Q_{e+p}/h_{Poyn}$ ratios. However, as pointed out in \citet{Rap07}, Eq.~10 should be regarded only as a lower limit of the Parker angle, as it does not take into account the current sheet formation.  On the other hand, much higher values of $\theta_D$ are predicted in \citet{Hen96, Gal96}.

\section{Average thick target X-ray spectrum from nanoflares}
In this section we attempt to model the expected form of the thick target spectrum produced by the electrons in the active region loops. We first calculate the distribution of the accelerated electrons' kinetic energy.
The total number of electrons, $N_e$, produced by each loop is given by the product of the loop electron density $n_e$ (given by \citet{Ros78}) multiplied by the plasma volume $V_{rec}$ entering the current sheet during
the reconnection time $t_{rec}$. Therefore,  $N_e = 2 v_{inflow} t_{rec} n_e A_{rec}$. We assume that the kinetic energy distribution in each loop lies in the energy range $\Delta E = [ E_{kin\, (min)} , E_{kin\, (max)} ] $,
where $E_{kin\, (min)}$ and $E_{kin\, (max)}$ are the energy limits calculated by Equation~(\ref{kinetic}). The amplitude of the kinetic energy distribution of each loop is equal to $g=N_e/\Delta E$ and is assumed to be constant inside $\Delta E$.
To compute the kinetic energy distribution of all selected loops, we divide the energy E from 0.1~keV to 7~keV in 100 energy bins of $\delta E=0.07$~keV. For each energy bin, at a given energy E, we take the sum of the individual loop 
distribution amplitudes $g_j$ for which ($E_{kin\, (min)\, j}<E<E_{kin\, (max)\, j}$). Also, for each energy bin we take the sum of the volumes $V_{loop\, j}$ of the corresponding loops. Finally, the average kinetic energy 
distribution $F(E) dE$ is calculated at each energy bin $i$ as $F(E_i) = \Sigma g_j/\Sigma V_{loop\, j}$, in units of (electrons cm$^{-3}$ keV$^{-1}$). Note that this distribution assumes that nanoflares from all
loops are triggered simultaneously, which is one more simplification. Our derived active region kinetic energy distribution is shown in Figure~\ref{x-ray_flux:fig}a. We performed a power law fit of the form $F= G E^b$  and found an exponent $b \simeq\, -4$ 
and a proportionality factor $G =3.3\times 10^9$~cm$^{-3}$ keV$^{-1}$. We used the derived fit parameters to compute the X-ray spectrum assuming the thick target approximation \citep{Br71} using the Fortran code developed by 
Holman \citep{Holm01}. We assumed that the area of the radiating source function equals the sum of the loop footpoints areas which is $A=7 \times 10^{19}$~cm$^2$. The resulting X-ray spectrum is seen in Figure~\ref{x-ray_flux:fig}b.
The X-ray spectrum is divided by the number of loops (5000) assuming that on average only one nanoflare is active at a time. Therefore, the computed X-ray represents a lower limit of the nanoflare emission. The computed X-ray flux per 
loop has a maximum at 3~keV, of value $10^{-2}$ photons s$^{-1}$ keV$^{-1}$ cm$^{-2}$, and falls at $10^{-4}$~photons s$^{-1}$ keV$^{-1}$ cm$^{-2}$, at 7~keV. 
Our derived X-ray spectrum exhibits a narrow energy range because of the narrow energy range of the electron kinetic energies. For this reason, a comparison of the spectrum shape with low solar activity X-ray observed spectra, as the RHESSI data presented in \citet{McT09}, or the SphinX data present in \citet{Mice12} was not attempted. However, our computed X-ray flux is of the same order of magnitude as
the upper limits measured  in the quiet Sun with RHESSI, (\citet{Han11}, their Figure~12, left panel, page 278). 

\section{Loop hydrodynamic response to nanoflares}
\label{section_hydro}

In this section, we employ the heating rates computed via Eq.~\ref{heating} as an input to hydrodynamic loop simulations.
This allows to determine the thermal response (differential emission measure; DEM) to the heating resulting from our model. 
As the DEM can be deduced also by observations, comparison between simulated and observed DEMs is a standard test-bed for coronal models.

We assume that each loop is heated due to the activation of several 
current sheets of sub-telescopic sizes, which inject beams of 
accelerated particles into the loops. Each current sheet is activated for a 
duration of the order of the Alfv\'en crossing time along the loop, 
but the nanoflare cascade duration is different and subject to 
parametrization.

We also assume that after the particles (electrons and protons)
are accelerated somewhere along the loop, they exit the current sheet,
and they deposit their kinetic energy, via Coulomb collisions, to the 
lower and denser parts of the loop (i.e., thick-target model of electron 
beams; e.g. \citet{Br71}). However, the general characteristics (e.g. 
maximum temperature and density), of loops submitted to thermal 
(i.e., direct) and non-thermal  (i.e., particle) heating do not 
substantially differ for the same total energy release (e.g. 
\citet{Wa04}). Differences could arise in the ultra-hot plasma 
($> 5$~MK), during the early stages of impulsive heating (e.g. 
\citet{Klim08}) or in the mass flows from the loop footpoints, 
which are however not considered here. Instead, particle heating is treated here
as thermal heating. Finally the particle beams follow the magnetic field lines directly 
connected to their individual current sheet whereas the macroscopic 
loop is heated due to the collective effect of all individual current 
sheets.

In order to study the loop plasma response to the calculated heating 
rates we use the EBTEL model of \citet{Klim08}. EBTEL is a 0D model, 
i.e. it considers spatially-averaged properties, and uses analytical 
approximations to solve the time-dependent  hydrodynamic equations.
It was found in good agreement with simulations using far more complex, 
yet computationally expensive, 1D models. By definition then, 
the heating is assumed to be spatially uniform in EBTEL. Allowing 
for different scenarios of the spatial localization of the heating
leaves distinct signatures only during the early stages of the 
hydrodynamic evolution, when the loop is at very high temperatures 
(e.g. \citet{Pats05}). For a given heating profile, loop length, 
and initial temperature and density conditions,  EBTEL calculates
at any instance the  coronal temperature and density as well as the 
transition region (footpoint) and coronal DEM. 
The application of our model to the observations of NOAA 09114 described
in the previous paragraphs supply to each loop hydrodynamic simulation the 
length and the corresponding heating rate.
Given that the model is 0D
and analytical, it can calculate numerous solutions within feasible computer
times.

Our hydrodynamic calculation starts with initial conditions determined according to the scaling law 
of to \citet{Ros78}, at a coronal temperature of 1~MK. (However note that this scaling law is derived in the 
hydrostatic limit. Foe the use of this approximation see our Section~9).
Then, and corresponding to t=0 s, each loop was submitted to impulsive
heating, given by Equation~\ref{heating}, so that each loop is heated 
impulsively by the corresponding heating rate of Figure~\ref{heat_functions:fig}.
The heating took the form
of a step function, with a duration of $t_{heat}$. 
Numerical simulations \citep{Geor98} predict a wide distribution for the duration of heating events in nanoflares. Therefore, $t_{heat}$ is a free parameter for our model, and we run simulations 
with the values $t_{heat}=$15~s, 50~s, 100~s, 250~s, and 500~s, for all the loops. We also considered the case of 
$t_{heat}$ set equal to the Alfv\'en crossing time for each loop. Since $t_{heat}$ is in general longer than the
Alfv\'en crossing time, this computation can correspond to a storm of nanoflares when the different fragments of 
the loop current sheet are activated at different times.
The employed $t_{heat}$ values are compatible with the duration of small-scale impulsive
energy release events found in MHD simulations of coronal heating. 
For each loop the corresponding simulations lasted for 5000s to allow
both the heating and cooling of plasma to be followed. 

\begin{figure}[!h]
\epsscale{.80}
\plotone{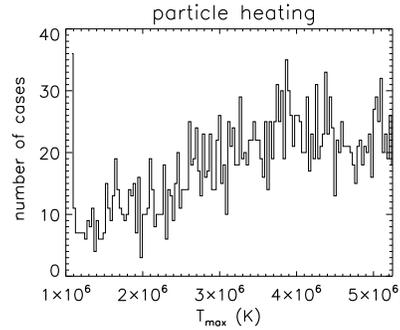}
\caption{Histogram of temperatures achieved at the maximum of each loop time-averaged DEM distribution. This calculation is performed for 
$t_{heat}=100$~s. }
\label{tmax_figure}
\end{figure}

In Figure~\ref{tmax_figure}, we plot the histogram of the 
temperatures of the peak of the temporally-averaged DEM of each loop computed for $t_{heat}$=100~s. In Fig.~\ref{tmax_figure}, the vertical axis represents the number of loops per temperature bin.
We can see that the particle heating creates an almost uniform distribution of temperatures. The peak of the distribution at 10$^6$~K corresponds to loops not significantly influenced 
by the heating and keeping their initial temperature. For $t_{heat}=50$~s the temperatures histogram is very similar to the one of Fig.~\ref{tmax_figure}. 
For $t_{heat}=500$~s, the distribution exhibits a plateau in the range of 2~MK up to 7~MK, while for $t_{heat}=15$~s, the histogram shows a higher probability of temperatures lower than 2~MK.

\begin{figure}[!h]
\epsscale{.8}
\plotone{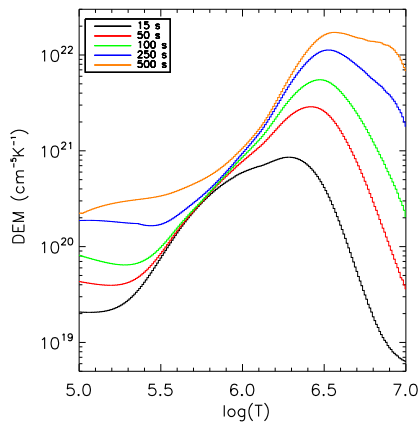}
\caption{Differential Emission measure calculated with the heating 
caused by particle acceleration calculated for different $t_{heat}$.}\label{dem_figure}
\end{figure}

In Figure~\ref{dem_figure}  we plot the time-averaged DEMs for all the 
considered loops in NOAA AR~09114. The plotted DEMs  correspond to 
both the coronal and footpoint parts of the modeled loops. This is legitimate
because we deal with averages over the entire AR which would 
obviously include contributions from both the coronal and footpoint regions. 
The latter are known to supply, particularly in AR cores, most of the low 
temperature emission at around 1~MK.

DEMs provide the amount of plasma present at each temperature bin and 
therefore offer an idea of the thermal distribution of the region or feature 
in question. Several remarks are now in order from Figure~\ref{dem_figure}.
The deduced DEMs has maximum values $9\times 10^{20}$, $2\times 10^{21}$, $4\times 10^{21}$, 
$10^{22}$ and $1.5\times 10^{22}$~$\mathrm{cm^{-5}\, K^{-1}}$ for $t_{heat}$ of 15, 50, 100, 250, and 500~s respectively. 
The temperature of the maximum DEM value raises for higher heating durations and it is found in the range of 6.4 to 6.6. in log(T).

Observationally deduced DEMs  (e.g. \citet{La97}; \citet{ODwy11}:
their Figures 4 and 14 respectively) exhibit a broad peak from 
~ 6-6.5 in log of T at a value of few times $10^{21} \mathrm{cm^{-5}\, 
K^{-1}}$. At both limits of this plateau, DEMs drop off rapidly.
Therefore, the DEMs from our model can reproduce the 
features of observed AR DEMs. 
Once again we emphasize that, in the framework of our
proof-of-concept calculations, we are not aiming to reproduce
any particular observational detail.

\section{Discussion}
Since a rather large number of assumptions and/or approximations were introduced in our modeling of nanoflare heating presented in the previous sections, we 
summarize here the main limitations and conditions of validity of our model, as well as possible future extensions.
\subsection{Model limitations}
An important limitation of our model is that it does not describe the initial
process of transformation of photospheric Poynting flux into magnetic free energy, which, when a certain critical 
value of the Parker angle is reached, is presumably released back into the plasma causing heating. Here, we simply assumed that all loops reach the critical point (Parker angle) at which the Poynting flux is transformed into particle acceleration. 

It should be noted that reconnection is a complex non-steady phenomenon \citep{Lou07, Sam09}. The presence of a guide magnetic field component has an important 
influence yet not fully understood \citep{Ya10, Birn07}. Another important approximation is the use of a Harris type analytical geometry to study the orbits of accelerated particles. Such an approach does not 
take into account either the perturbations caused by the accelerated particles onto the fields or the more complex structures of the magnetic field topology that we expect to be formed at the reconnection sites.

The scaling laws  of \citet{Ros78}  were presently used in order to compute initial conditions for our nanoflare simulations, as well as for the 
calculation of kinetic energies and heating rates in the previous sections. 
However, these formulae are valid in a strict sence for hydrostatic atmospheres, while in reality all parameters in our calculations should exhibit some time dependence. 
One may remark, nevertheless, that the characteristic timescale of evolution of the atmosphere is determined by the 
time $t_{evap}$ needed by the chromospheric evaporation flows to fill the loops with dense and hot plasma. According to some large flare simulations (\citet{Yo98}),  chromospheric evaporation flows at 0.2-0.3 of the sound speed, at a temperature $\simeq\ 4\,\times\, 10^6$~K. We find $t_{evap}\,=\, L/(0.4\, v_s)$ in the range of 100~s to 500~s for 80\%\ of our cases. The resulting time-scales are of the same order with the highest $t_{heat}$ values used in our calculations.
At any rate, \citet{Yo98} argue that the chromospheric evaporation flow should not influence the reconnection rate at a flare's X-point. In view of the above, and since $t_{evap}$ is typically larger or at most equal to $t_{heat}$, we conclude 
that the use of \citet{Ros78} scaling laws in our heating computations is an allowable approximation. A more accurate computation would require a proper application of a chromospheric evaporation particle heating rate feedback model.
This is proposed for future work.

\subsection{Comparison with other models}
Some words are necessary to explain our choice of hydrodynamic model. 
There are basically two approaches to study coronal loop heating based on i) 3D MHD  (e.g.,\citet{Gud05}; \citet{Pet06}; \citet{Dah12};\citet{Bing11}), or ii) 1D and 0D hydrodynamic simulations. 3D MHD simulations supply a physics-based heating function (e.g., Ohmic
heating at intense current sheets formed at the interfaces of braided magnetic elements). However 3D MHD simulations lack
the spatial resolution available in 1D hydrodynamic simulations. A high resolution, on the other hand, is crucial for an accurate description of the plasma thermodynamic 
response to a given heating. Nevertheless, the heating functions selected in 1D hydrodynamic loop simulations is ad-hoc, i.e., they can be chosen arbitrarily. Finally, hydrodynamic descriptions are far less computationally expensive than 3D MHD and thus more appropriate for extensive studies of thousands of loops.

A future improvement could concern the geometry of our model. Replacing the Harris current sheet with a more realistic geometry would allow to calculate test particle orbits for a range of 
parameters pertinent to solar active regions, over a large number of coronal loops. Of course, the ultimate improvement would be to simulate the feedback of the plasma response due to the chromospheric evaporation on the acceleration of the particles.

\section{Conclusions}
 
In the present study, we provide a set of calculations for nanoflare 
heating in coronal loops based on a composite model in which 
the heating term used for hydrodynamic simulations of 
nanoflares is provided by considering particle acceleration 
in reconnecting current sheets. Our main steps and conclusions 
are the following:

1) Our calculations are utilizing the data of observations: (a) The general structure of the magnetic field 
is deduced by means of a current-free
(potential) magnetic field extrapolation of an observed active 
region's (NOAA AR 9114) magnetogram. (b) We selected 5000 closed magnetic field lines, derived from the extrapolation,
to represent coronal loops in which we will study the nanoflares. 
(c) Poynting flux is supplied in current sheets, one for each coronal loop, 
produced by photospheric motions at the loop footpoints. The Poynting flux 
is calculated using the measured magnetic fields and the estimated values for the inductive velocities at the photospheric level.

2) In our current sheets, reconnection always occurs, because we assume that the discontinuity in the magnetic field configuration has reached 
a critical mis-alignment angle. The mis-alignment angle, $\theta_D$, (twice the 
adopted Parker angle) varies randomly from loop to loop in a uniform distribution from $8\degr$ to $50\degr$.
In this model we compute the physical conditions in the current sheets.
The induced electric field $E_{rec}$ is in the range 0.01 to 100~V/m, and is larger than the Dreicer electic field which favors the direct acceleration of particles.
The plasma inflow velocity $v_{inflow}$ is in the range of 0.1 to 100~\kms. 

3) The Poynting flux supplied by the photospheric motion is entirely transformed into kinetic energy of the particles, accelerated 
in the reconnecting current sheets. The final kinetic energies of electrons and protons are calculated using analytic formula derived 
in test particle studies \citep{Efth05,Litv96}. The electron kinetic energy gain turn to be up to  8~keV, while for protons it turn to be in the range of 0.3 to 470~keV.

4) We consider the process of particles' acceleration as the unique source of plasma heating. This assumption is supported by the fact that, 
at least in large flares, electron acceleration corresponds to 50\% of the released energy \citep{Birn07}. 
We use a simple phenomenological expression (Eq.~\ref{heating}) to compute the heating rate produced by accelerated electrons $Q_e$ and protons $Q_p$. 
The produced heating rates are in the range of $10^{-4}$ to 1 $\mathrm{erg\, s^{-1} cm^{-3}}$ while the ratio $Q_p/Q_e$ takes values in the range 0.1 to 5 and is 
higher than 1 in 50\% of cases. 

The power law of the computed heating rate as a function of the loop length, derived both via a fit to the calculated data and via an analytical derivation
yields an exponent of $\simeq -1.5$, which falls within the constrains derived from observations \citep{Man00}. Moreover, we found a linear dependence of the
heating functions on the Poynting flux at the footpoints and a trigonometric dependence on the angle $\theta_D$.

5) We computed the form of X-ray spectra generated by the accelerated electrons from all loops, using the \lq thick target\rq\ approach. The derived spectrum has a peak intensity of 
$10^{-2}$~photons s$^{-1}$ keV$^{-1}$ cm$^{-2}$ at 3~keV and decreases with a power law shape and an exponant equal to $\simeq -4$. This result is in agreement with 
today upper limits derived from observations \citep{Han11}. 

6) Finally we performed hydrodynamic simulations using the 0D EBTEL code to compute the characteristic atmospheres of our loops. The constrains of the simulations are the derived heating rates and the 
loop length while the heating event duration is kept as a free parameter. The deduced DEMs have maximum values in the range of $9\times 10^{20}$ to $1.5 \times\, 10^{22}$~$\mathrm{cm^{-5}\, K^{-1}}$
for temperatures from 6.4 to 6.6 in log(T). These derived values are in agreement with DEMs derived from observations \citep{La97,ODwy11}. 

7) We discuss the various limitations of our model and we propose a number of possible future extensions as well as comparisons with other models in the literature.

{\bf Acknowledgements:} 
{\bf We thank the anonymous referee for helpful and constructive comments.}
This research has been supported in part by the European Union (European 
Social Fund – ESF) and in part by the Greek Operational Program 
"Education and Lifelong Learning" of the National Strategic 
Reference Framework (NSRF) - Research Funding Program: Thales. 
"Hellenic National Network for Space Weather Research"-MIS 377274.
S.P. acknowledges support from an FP7 Marie Curie Grant (FP7-PEOPLE-2010-RG/268288).
C.G. acknowledges support from program 200/790 of the research committee of the Academy of Athens.

\end{document}